\documentclass[prl,twocolumn,english,superscriptaddress,footinbib]{revtex4-1}

\usepackage{amsmath,amssymb,mathrsfs}
\usepackage{amsfonts,bm}
\usepackage{amsmath}
\usepackage{amssymb}
\usepackage{amsthm}
\usepackage{graphicx}
\usepackage{graphics}
\usepackage{subfigure}
\usepackage{color}
\usepackage{bm}
\usepackage{hyperref}
\usepackage{rotating}
\usepackage{comment}
\usepackage[colorinlistoftodos]{todonotes}

\usepackage{xfrac}

\newcommand{\be}{\begin{equation} }
\newcommand{\ee}{\end{equation} }
\newcommand{\ba}{\begin{eqnarray} }
\newcommand{\ea}{\end{eqnarray} }

\newcommand{\bpm}{\begin{pmatrix}}
\newcommand{\epm}{\end{pmatrix}}
\newcommand{\bmm}{\begin{matrix}}
\newcommand{\emm}{\end{matrix}}

\newcommand{\la}{\label}
\newcommand{\p}{\partial}
\newcommand{\bea}{\begin{eqnarray}}
\newcommand{\eea}{\end{eqnarray}}

\makeatother

\usepackage{babel}

\begin{document}

\title{ Hydrodynamics of two-dimensional compressible fluid with broken parity: 
\\
variational principle and free surface dynamics in the absence of dissipation}
 \author{Alexander G.~Abanov}
\affiliation{Simons Center for Geometry and Physics, Stony Brook, NY 11794, USA}
\affiliation{Department of Physics and Astronomy, Stony Brook University, Stony Brook, NY 11794, USA}
\author{Tankut Can}
 \affiliation{Initiative for the Theoretical Sciences, The Graduate Center, CUNY, 10012, USA}
 \author{Sriram Ganeshan}
 \affiliation{Department of Physics, City College, City University of New York, New York, NY 10031, USA }
 \author{Gustavo M. Monteiro}
 \affiliation{Instituto de F\'isica Gleb Wataghin, Universidade Estadual de Campinas-UNICAMP, 13083-859 Campinas, SP, Brazil}
 \affiliation{International Institute of Physics, Campus Universit\'ario Lagoa Nova, Natal RN, 59078-970, Brazil}

\date{\today}

\begin{abstract}
 We consider an isotropic compressible non-dissipative fluid with broken parity subject to free surface boundary conditions in two spatial dimensions. The hydrodynamic equations describing the bulk dynamics of the fluid as well as the free surface boundary conditions depend explicitly on the parity breaking non-dissipative odd viscosity term. We construct a variational principle in the form of an effective action which gives both bulk hydrodynamic equations and free surface boundary conditions. The free surface boundary conditions require an additional boundary term in the action which resembles a $1+1D$ chiral boson field coupled to the background geometry. We solve the linearized hydrodynamic equations for the deep water case and derive the dispersion of chiral surface waves.  We show that in the long wavelength limit the flow profile exhibits an oscillating vortical boundary layer near the free surface. The thickness of the layer is controlled by the length scale given by the ratio of odd viscosity to the sound velocity $\delta \sim \nu_o/c_s$. In the incompressible limit, $c_s\to \infty$ the vortical boundary layer becomes singular with the vorticity within the layer diverging as $\omega \sim c_s$. The boundary layer is formed by odd viscosity coupling the divergence of velocity $\bm\nabla \cdot \bm{v}$ to vorticity $\bm\nabla \times \bm{v}$. It results in non-trivial chiral free surface dynamics even in the absence of external forces. The structure of the odd viscosity induced boundary layer is very different from the conventional free surface boundary layer associated with dissipative shear viscosity. 

\end{abstract}
\maketitle

\textit{\textbf{Introduction.}}  In a seminal work,  Avron ~\cite{avron1998odd} noticed that in two spatial dimensions the viscosity tensor allows for a parity breaking term, dubbed \emph{odd viscosity}, without breaking the fluid isotropy. The recent interest in this parity violating stress-shear response was triggered by Ref.~\cite{avron1995viscosity}, which shows that the odd viscosity coefficient is quantized for quantum Hall systems, where it is related to the adiabatic curvature on the space of flat background metrics on a torus. 
Several theoretical works subsequently studied odd viscosity, also known as \emph{Hall viscosity}, as a new quantized observable of quantum Hall fluids~\cite{tokatly2006magnetoelasticity,tokatly2007new,tokatly2009erratum, read2009non,haldane2011geometrical,haldane2011self,hoyos2012hall, bradlyn2012kubo, yang2012band,abanov2013effective,hughes2013torsional, hoyos2014hall, laskin2015collective, can2014fractional,can2015geometry,klevtsov2015geometric,klevtsov2015quantum, gromov2014density, gromov2015framing, gromov2016boundary, scaffidi2017hydrodynamic, andrey2017transport,alekseev2016negative,pellegrino2017nonlocal}.  Other classes of fluid systems where `odd viscous' effects might be important include polyatomic gases~\cite{korving1966transverse, knaap1967heat, korving1967influence, hulsman1970transverse}, chiral active matter \cite{banerjee2017odd, souslov2019topological, soni2018free}, vortex dynamics in two dimensions~\cite{wiegmann2014anomalous, yu2017emergent, bogatskiy2018edge, bogatskiy2019vortex}, chiral superfluids/superconductors~\cite{read2009non, hoyos2014effective} and any fluid dynamics with parity breaking intrinsic angular momentum of constituent particles. Experimentally, odd viscosity can be indirectly measured through corrections to the charge transport (see, e.g., Ref.~\cite{berdyugin2019measuring}) or more directly in chiral active fluids by observing the dynamics of the boundary of a fluid with odd viscosity~\cite{soni2018free}. It was also shown that if the fluid is almost incompressible, the odd viscosity $\nu_o$ effects are most visible at the dynamical boundary subject to no-stress (free surface) boundary conditions~\cite{ganeshan2017odd, abanov2018odd}. 
 
A classic example of no-stress dynamical boundary conditions is that of surface gravity waves. For perfect fluids without any viscosities, the irrotational surface dynamics was solved by Stokes in 1847 \cite{george1847stokes}, when he derived the famous surface gravity wave dispersion relation $\Omega=\pm \sqrt{g |k|}$ \footnote{This dispersion is an approximation to deep water waves.}. 

The presence of infinitesimal shear viscosity requires tangent stress induced vorticity near the moving boundary thereby violating irrotationality. This effect can be captured within linearized hydrodynamics, where a small but finite shear viscosity creates a thin vortical boundary layer of thickness $\delta \sim \sqrt{\nu_e /\Omega}$ near the surface of the fluid~\cite{lamb1932hydrodynamics}. The fluid outside this layer remains irrotational to an excellent approximation~\footnote{Both the fluid velocity and vorticity inside the layer are finite in the limit $\nu_e\to 0$ producing small damping of the surface waves vanishing in the limit of an ideal fluid $\nu_e\to 0$. The damping rate is given by a well known result quadratic in the wave vector $2\nu_e k^2$~\cite{lamb1932hydrodynamics}.}. The presence of odd viscosity in addition to shear viscosity significantly alters this boundary layer structure as shown by some of us in Ref.~\cite{abanov2018odd}. While the thickness of the boundary layer is still controlled by $\delta \sim \sqrt{\nu_e/\Omega}$, in the limit $\nu_e\rightarrow 0$ (with fixed $\nu_o$), the vorticity $\omega =\bm\nabla\times\bm{v}$ within the layer diverges as $\omega\sim \frac{1}{\sqrt{\nu_e}}$~\cite{abanov2018odd}. Therefore, the limit $\nu_e\to 0$ is singular and can be thought of as the formation of a discontinuity in the velocity component tangential to the fluid boundary. This discontinuity makes it difficult to access free surface results for fluids with exactly zero shear viscosity $\nu_e=0$ (to be contrasted with the limit $\nu_e\to 0$ \cite{abanov2018odd}).

Strictly non-dissipative fluids with $\nu_e = 0$ are of considerable interest, with notable examples being quantum Hall fluids and superfluids. Therefore, in this work we consider a strictly non-dissipative fluid with odd viscosity and free boundary. The dissipationless nature of such a fluid allows for a variational principle description, which produces both bulk hydrodynamic equations and appropriate free surface boundary conditions. 

The solution to the linearized hydrodynamic equations with free surface shows that a \emph{finite compressibility} is needed to satisfy the no-stress boundary conditions at the free surface without appealing to weak solutions with discontinuities. The incompressible limit is therefore subtle and characterized by the formation of a singular boundary layer whose thickness is controlled by a new length scale $\delta \sim \nu_o/c_s$ (ratio of odd viscosity to sound velocity) with diverging vorticity ($\omega \sim c_s$).

\begin{figure}
\centering
\includegraphics[scale=0.3]{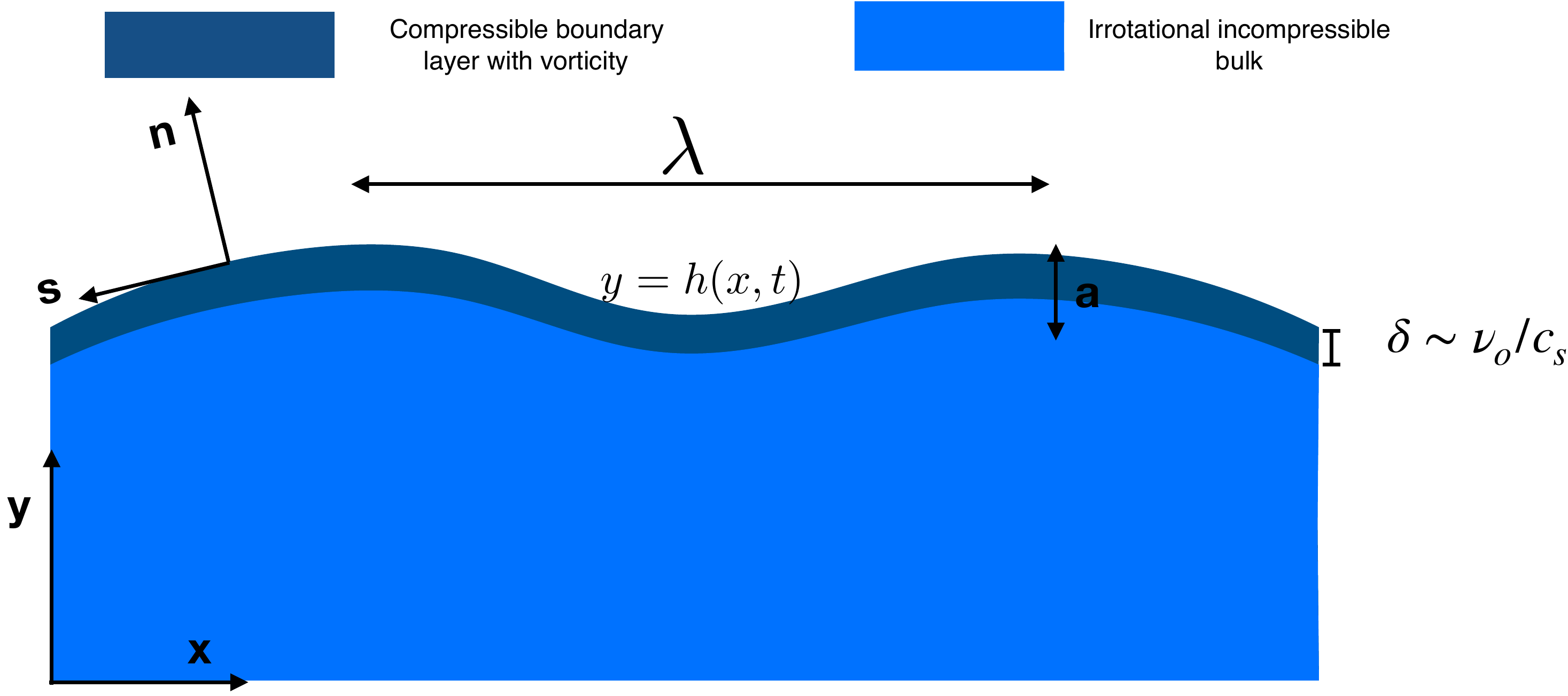}
\caption{Schematic of the free surface dynamics of an infinitely deep two-dimensional ``ocean''.}
\label{fig:schematic}
\end{figure}  

\textit{\textbf{Hydrodynamics of compressible fluid with odd viscosity.}} Hydrodynamic equations consist of the conservation equations for local mass and momentum \cite{footnote-energy}, assuming all other relevant quantities are equilibrated. The momentum conservation and continuity equation can be written in terms of the mass density of the fluid $\rho$ and its velocity $v_i$ as,
\begin{align}
	\p_t\rho+\p_i(\rho v_i)=0\,,	\;  \p_t (\rho v_i)+\p_j(\rho v_iv_j-T_{ij})=\rho F_i\,.
 \label{eq:eom}
\end{align}
Here we assumed that the fluid is charged so that the external force in the presence of electromagnetic fields is given by $F_i=-e(B \epsilon_{ij}v_j+E_i)/m$ with $\epsilon_{ij}$ being the Levi-Civita tensor. Additionally, we assume that the ratio between charge and mass density is constant and proportional to $-e/m$.  In the following we set $e/m=1$. For simplicity, we neglect thermal effects in this paper and do not consider local energy conservation \cite{footnote-energy}, since it follows directly from the equations for momentum and mass conservation.

For a fluid with odd viscosity, the stress tensor $T_{ij}$ in Eq.~(\ref{eq:eom}) is given by \cite{avron1998odd}
\begin{align}
	T_{ij}=-p\,\delta_{ij}+\nu_o \rho(\p_i^*v_j+\p_iv_j^*) \,,
 \label{eq:Tij}
\end{align}
where $\nu_o$ is called \textit{kinematic odd viscosity}.  Here and in the following, we use the star operation defined by $a_i^*\equiv \epsilon_{ij}a_j$. The pressure $p$ in Eq.~(\ref{eq:Tij}) must be understood as a function of the density $p(\rho)$ \footnote{The fluid with $p=p(\rho)$ is known as barotropic fluid. More generally, one should include entropy density field $s$ and write $p=p(\rho,s)$. In this case the conservation of energy equation should be added to (\ref{eq:eom}).}. Both the parameter $\nu_o$ and the \textit{equation of state} $p(\rho)$ are supposed to be derived from an underlying microscopic model and are assumed to be known in the rest of the Letter. Formally, the incompressible limit can be achieved by taking the limit of infinite sound velocity $c_s\to \infty$, where $c_s = \sqrt{dp/d\rho}$.

Eqs.~(\ref{eq:eom}) and (\ref{eq:Tij}) are often referred to as the first-order hydrodynamics, emphasizing the fact that the gradient expansion of the stress tensor (\ref{eq:Tij}) is stopped at the first order in spatial derivatives of velocity. It is well known that in the hydrodynamics with gradient terms the velocity of the fluid is not uniquely defined \footnote{Changing the definition of velocity by gradients is known as \emph{hydrodynamic frame redefinition}.}. One could fix the definition of velocity by saying that the mass density current entering the continuity equation in (\ref{eq:eom}) is given by $\rho v_i$. Alternatively, one could insist on $\rho v_i$ being the momentum density entering the second equation in (\ref{eq:eom}). {\it In the following, we assume that for the fluid under consideration the momentum density and the mass current density are identical and given by $\rho v_i$}. This may not be true in a particular microscopically realized fluid (see, e.g., \cite{moroz2019bosonic}) and the results presented in this letter will change. However, it is straightforward to generalize our calculations to such cases. 

For a fluid domain with boundaries, we must supply boundary conditions to the bulk equations of motion in (\ref{eq:eom}). The fluid free surface is a dynamical interface $\Gamma$ between two fluids where we impose one kinematic and two dynamical boundary conditions
\begin{align}
	(\p_t\Gamma)_n=v_n\Big|_{\Gamma}\,, \qquad n_iT_{ij}\Big|_{\Gamma}=0\,.
 \la{eq:bcs}
\end{align}
Here $n_i \equiv (n_x, n_y)$ are the components of the unit vector outward normal to the free surface $\Gamma$ (see Figure~\ref{fig:schematic}). The equation on the left is the kinematic boundary condition which states that the velocity of the fluid normal to the boundary is equal to the speed of the boundary. The pair of equations on the right states that there are no normal and tangent forces acting on an element of the fluid surface. It is important to realize that both conditions depend on how we parametrize momentum density and mass current in terms of the velocity field. The normal velocity $v_n$ entering the first condition arises from the mass current density, while the dynamic boundary conditions are given in terms of the stress tensor components $T_{ij}$. The velocity dependence in Eq.~(\ref{eq:Tij}) comes from the identification of the fluid momentum density with $\rho v_i$. 

\textit{\textbf{Variational principle.}} 
The fluid dynamics described in Eqs.~(\ref{eq:eom}-\ref{eq:bcs}) is non-dissipative (see \cite{footnote-energy}). The absence of dissipation allows us to capture the full bulk and boundary dynamics in a variational principle. For that we parametrize the flow velocity in terms of three scalar fields $(\theta,\alpha,\gamma)$, known as Clebsch potentials \footnote{Although the fluid Hamiltonian is only a function of mass density and flow velocity, the Poisson algebra between these quantities is degenerate, due to the existence of Casimirs. To overcome this difficulty, the phase space must be enlarged. In the enlarged phase space, density and Clebsch potentials become canonical variables.} 
\be 
	v_i=u_i-\nu_o \p_i^*\ln\rho\,,
	\quad u_\mu\equiv\p_\mu\theta+A_\mu+\alpha\p_\mu\gamma\,,   
 \la{Clebsch}
\ee
where $i=1,2$ and $\mu=0,1,2$. For an introduction to Clebsch parametrization we refer to Ref.~\cite{1997-ZakharovKuznetsov} and references therein. 

Let the fluid domain $\mathcal M$ be given by $y\leq h(t,x)$ and let $\varepsilon(\rho)$ be the internal energy density of the fluid. Using the definition (\ref{Clebsch}), the hydrodynamic action can be written as
\begin{align}
	&S=S_{\mathcal M}+\,S_\Gamma\,, 
 \la{hydro-action} \\
	&S_{\mathcal M}=-\iint dt\, dx\int\limits_{-\infty}^{h(t,x)} dy
	\Big[\rho\left(u_0+\tfrac{1}{2}v_i^2\right)+\varepsilon(\rho)\Big]\,, 
 \la{action-bulk} \\
	&S_\Gamma=\nu_o\iint dt \,dx
	 \Big[2\phi_t  \sqrt{\tilde\rho(1+h_x^2)}-
	\tilde \rho\, h_x h_t -\phi_x\phi_t
	\Big]\,, 
 \la{action-edge}
\end{align}
where Eq.~(\ref{action-edge}) is defined on the boundary $y=h(x,t)$. The boundary value of the density $\tilde\rho$ is defined in terms of the bulk density as $\tilde\rho(x,t)= \rho (x,h(x,t),t)$. The auxiliary field $\phi(x,t)$ is restricted to the boundary and is necessary to guarantee the action invariance with respect to boundary reparametrizations. Note that the action for $\phi$ resembles the one for a chiral boson coupled to the boundary geometry. The boundary action (\ref{action-edge}) does not affect the hydro equations in the bulk, but is necessary to ensure the no-stress boundary conditions at the boundary. 

Variation of Eq.~(\ref{hydro-action}) with respect to $\theta$ gives both the continuity equation in the bulk and the kinematic boundary condition. Variation over potential $\gamma$ on the boundary also provides the kinematic boundary condition, whereas the variation of $\rho$ taken at the boundary relates the field $\phi_t$ with $h_x,\tilde\rho$, and the boundary value of $v_x$. Momentum conservation comes from the bulk equations of motion for $\rho,\theta,\alpha,\gamma$ together with Eq.~(\ref{Clebsch}) and the equation of state defining pressure as a function of density, that is, $p = \rho\, d\varepsilon/d\rho-\varepsilon$. Normal and tangent dynamical boundary conditions, Eq.~(\ref{eq:bcs}), arise from variations over $h$ and $\phi$, respectively. For details, see supplementary information \cite{SM}.

\textit{\textbf{Casimirs and Hall constraint in the bulk.}}  
 Let us consider the fluid domain to be the whole plane with the action given by Eq.~(\ref{action-bulk}) with the upper limit of integration $h\to +\infty$. From the time derivative part of the action $-\rho u_{0}=-\rho \p_{t}\theta -(\rho\alpha) \p_{t}\gamma$ we can immediately derive Poisson brackets between $\rho,\theta,\alpha,\gamma$ and then between $\rho,u_{i}$ \cite{SM}. It is clear that the Poisson algebra between $u_{i}$ and $\rho$ is not affected by the odd viscosity term. Therefore, the corresponding Poisson brackets between $\rho$ and $v_{i}$ can be obtained  through a field redefinition (\ref{Clebsch}).  It is known that the Poisson structure between $u_i$ and $\rho$ is degenerate and possesses an infinite number of Casimirs. Defining the quantity $Q\equiv(\p_iu_i^*-B)/\rho$, one can show that
\be
	I_{r} = \int d^{2}x\, \rho\, Q^{r}
 \la{eq:IN}
\ee
is a Casimir for any $r=0, 1, 2, \ldots $; that is, $I_{r}$'s have vanishing Poisson brackets with any function of $\rho$ and $v_{i}$ and are therefore conserved for any fluid Hamiltonian. In terms of $v_i$ and $\rho$, the quantity $Q$ is
\be
	Q=\frac{\omega-B+\nu_o\Delta\ln \rho}{\rho} \,.
 \la{eq:Q}
\ee
For the dynamics given by Eq.~(\ref{action-bulk}), one can show that $Q$ is transported along the flow, that is, $D_t Q=0$, where $D_t \equiv \p_t +v_j\p_j$ is the material derivative. In particular, if this quantity is initially constant $Q=\beta$, it will remain constant at all times at all points. This observation allows to consider a reduction of the hydrodynamics (\ref{eq:eom}) subject to the constraint $Q=\beta$. If the external magnetic field is constant, the fluid density fluctuations become gapped and one can recognize this constraint as the so-called ``Quantum Hall constraint''\cite{stone1990superfluid,abanov2013effective}. For the quantum Hall fluid the constant value $\beta = \nu^{-1}\frac{h}{m}$ is given in terms of Planck's constant $h$, particle mass $m$ and quantum Hall filling fraction $\nu$. This constraint was originally derived by M. Stone in the fractional quantum Hall context starting from Chern-Simons-Ginzburg-Landau theory \cite{stone1990superfluid} and then  generalized to include  odd viscosity in Ref.~\cite{abanov2013effective}. Imposing the constraint $Q=\beta$ makes all Casimirs proportional to the total number of particles in the system and can be understood as a Hamiltonian reduction of the fluid dynamics.

We leave the investigation of the surface dynamics with the quantum Hall constraint for future work and focus here on the dynamics of a fluid without external fields: $E_i=0$, $B=0$. We also assume finite thermodynamic compressibility $dp/d\rho =c_{s}^{2}<\infty$ so that the fluid is compressible and supports sound waves. 

\textit{\textbf{Bulk dynamics. Propagating waves.}} Let us now consider small perturbations around the homogeneous background given by $\rho=\rho_0$, $v_i=0$ propagating in the bulk of the fluid. We linearize the hydrodynamic equations (\ref{eq:eom}) in $n=(\rho-\rho_0)/\rho_0$ and $v_i$ obtaining
\begin{align}
	\p_t n =- \p_i v_i \,,\quad
	\p_t v_i = -c_{s}^{2} \p_i n  + \nu_{o} \Delta v_i^*\,,
\label{eq:lineareom0}
\end{align}
where we used $\p_ip=c_s^2\p_i\rho$ with the sound velocity $c_{s}$ considered to be a constant computed at $\rho=\rho_{0}$. For plane wave solutions $(n,v_i) \propto e^{i\bm{q}\cdot\bm{r}-i\Omega t}$ we obtain the dispersion of linear modes given by 
\begin{align}
	\Omega =0\,, \qquad \Omega_\pm=\pm\sqrt{c_s^2 q^2 + \nu_o^2 q^4} \,.
 \label{eq:disp-bulk}
\end{align}
The corresponding (unnormalized) eigenvectors are 
\begin{align}
	(n,v_i)_0 =(\nu_o q^2, i c_s^2q_i^*),\quad
 	(n,v_i)_\pm =(q^2, \Omega_\pm q_i-i\nu_o q^2 q_i^*)
 \label{eq:nvi}
\end{align}
for $\Omega=0$ and $\Omega=\Omega_\pm$, respectively.

Let us denote the divergence of the fluid velocity as $\mathcal{D}=\p_i v_i$. This quantity is identically zero for incompressible flows. It is often convenient to use the divergence $\mathcal{D}$ together with vorticity $\omega$ instead of velocity components. In particular, the eigenmodes considered above can be written as $(n,\omega,\mathcal{D})_0 = q^2 (\nu_o,c_s^2,0)$ and $(n,\omega,\mathcal{D})_\pm = q^2 (1,-\nu_o q^2,\Omega_\pm)$. One can see that the first mode corresponds to linearly static perturbation, which is incompressible ($\mathcal{D} = 0$) with vorticity and density perturbations proportional to each other. For the $\pm$ modes the ratio $\omega/\mathcal{D}$ gives the ``tilt'' (slope) between the direction of the wave vector and velocity of the perturbation equal $\nu_o q^2/\Omega_\pm$. This tilt was discussed by Avron \cite{avron1998odd}.

It is clear from (\ref{eq:disp-bulk}) that the ratio $k_0=c_s/\nu_o$ defines a characteristic scale for the wave vector. For $q\ll k_0$ the corrections to bulk waves are small while for $q\gg k_0$ the odd viscosity effects are dominant and the sound waves become almost purely transversal.

\textit{\textbf{Linear surface dynamics.}} Let us now assume that the fluid is confined to a half-plane with perturbed boundary $y\leq h(x,t)$ and let us seek linearized solutions describing time evolution of the boundary $h(x,t)$ together with corresponding bulk and boundary density and velocity profiles. Focusing on velocity and density perturbations confined to the fluid boundary, we must look for solutions of the type $e^{m y -i kx -i\Omega t}$, with $\operatorname{Re}(m)>0$ defining the rate of the decay of the boundary perturbations into the bulk $y\to -\infty$. These boundary waves can be obtained from the bulk propagating solutions obtained in the previous paragraph through the substitution $(q_x,q_y,q^2)\to (k,-im,k^2-m^2)$. In particular, we obtain the following dispersion relation 
\begin{align}
	\Omega^2 = \nu_o^2(m^2-k^2)^2 - c_s^2 (m^2-k^2)\,.
 \la{eq:dispm}
\end{align}
This dispersion relation is real if either $m^2\leq k^2$ or $m^2\geq k^2 +k_0^2$. In the linearized regime, the general solution for boundary waves is a superposition of waves with given $\Omega$ and $k$ for two different values of $m$ \footnote{For given $k$ and $\Omega$, Eq. (\ref{eq:dispm}) is a fourth order polynomial for $m$ with four roots, only two of which have positive real part for $m^2\leq k^2$ or $m^2\geq k^2 +k_0^2$.}. Boundary conditions (\ref{eq:bcs}) fix the relative amplitude $A_1/A_2$ of the superposition of these modes. In the linear approximation, the boundary conditions (\ref{eq:bcs}) become:
\begin{align}
	h_t &= v_y\,,
 \label{eq:ht} \\
 	c_s^2 n +\nu_o(\p_x v_y+\p_y v_x) &= 0\,,
 \label{eq:Tnnbc}\\
 	\nu_o (\p_x v_x-\p_y v_y) &=0\,,
 \label{eq:Tsnbc}
\end{align}
all evaluated at $y = 0$. The equation (\ref{eq:ht}) determines the evolution of the surface if the normal component of velocity at the surface is known. For plane wave solutions with wave vector $q_{x} = k$ and frequency $\Omega$, the two remaining equations (\ref{eq:Tnnbc},\ref{eq:Tsnbc}) become
\begin{align}
 	 \nu_o(i k v_x -\p_y v_y) = 0\,,\quad
 	\nu_o (\p_y v_x+i kv_y) = -c_s^2 n\,.
 \la{eq:112}
\end{align}
This is an overdetermined system of equations for the dispersion if only a single mode at a given $m$ is used. Following Lamb \cite{lamb1932hydrodynamics} we look for the solution as a linear superposition of two boundary waves
\begin{align}
	(n,v_x,v_y) &= e^{ikx-i\Omega t} \sum^2_{\alpha=1}
	A_\alpha e^{m_\alpha y}\; (n^{\alpha},v_{x}^{\alpha},v_{y}^{\alpha})\,,
 \la{eq:twomodes}\\
 	 (n^{\alpha},v_{x}^{\alpha},v_{y}^{\alpha})&=\left(1, \frac{\Omega k}{k^2-m_\alpha^2} - \nu_o m_\alpha, 
	-i\frac{\Omega m_\alpha}{k^2-m_\alpha^2} +i\nu_o k \right) \,,
 \nonumber
\end{align}
where both values $m_{1,2}$ solve (\ref{eq:dispm}), and $(n^{\alpha},v_{x}^{\alpha},v_{y}^{\alpha})$ is a solution to the linearized bulk equations, and follows directly from Eq.~(\ref{eq:nvi}) by replacing $\bm{q}\to (k,-im)$. Substituting (\ref{eq:twomodes}) into (\ref{eq:112}) we obtain $A_1/A_2$ and the dispersion $\Omega(k)$. The incompressible regime is accessed when $k\ll k_0$.  In the following we focus on this regime, leaving the case of large $k$ for the supplementary \cite{SM}. 

Generally, in the presence of a confining potential (e.g. gravity) one expects to have both right and left propagating boundary modes. However, here we consider the surface in the absence of an external restoring force. In this case, one of the modes has $\Omega=0$, corresponding to an arbitrary initial profile $h(x)$ and zero initial velocity. The other mode is non-trivial and exists because odd viscous terms can play the role of a restoring force \cite{abanov2018odd}. The full equation for the dispersion of this mode is complicated. In the limit $k\ll k_0$, we have from (\ref{eq:dispm}) \cite{SM}
\begin{align}
	m_1 \approx k\left(1-\frac{(\Omega/\nu_o)^2}{2k_0^2 k^2}\right), 
	\quad
	m_2 \approx k_0\left(1+\frac{k^2}{2k_0^2}\right),
\end{align}
and the dispersion of the gapless mode 
\begin{align}
	\Omega \approx - 2 \nu_o k |k| + 2 \frac{\nu_o}{k_0}k^3+O(k_0^{-2}) \,. 
 \la{dispersion}
\end{align}
\begin{figure}
\centering
\includegraphics[width=8.5cm, height=3.5cm]{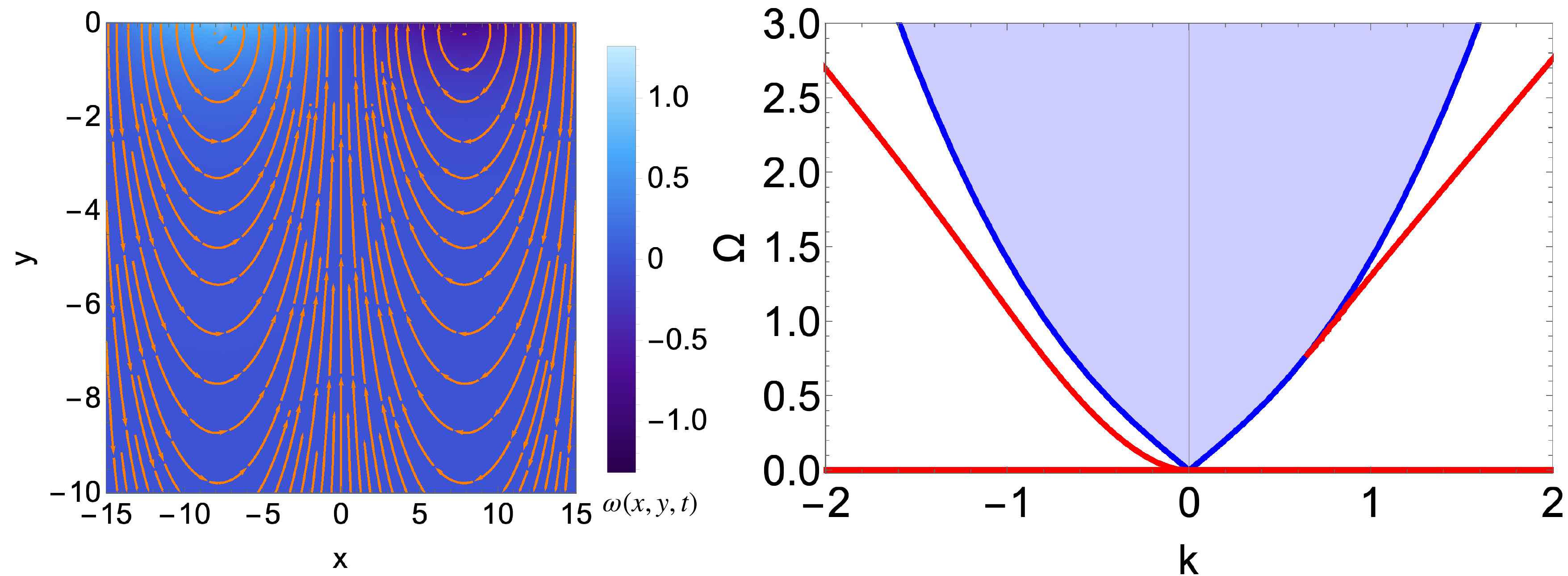}\\
\caption{[Right] Bulk (blue) and boundary (red) dispersion relation. [Left] Velocity profile vector field $(v_x(x,y), v_y(x,y))$ is plotted as a streamline plot with the color density denoting the vorticity profile $\omega(x,y)$ for time $t=0$ and $k=0.20551$.}
\label{fig:disp}
\end{figure}  
The leading term of the surface wave dispersion (\ref{dispersion}) is universal and coincides with the one obtained in Ref.~\cite{abanov2018odd}. However, in contrast with Ref.~\cite{abanov2018odd} the sub-leading in $k/k_0$ terms in (\ref{dispersion}) are also non-dissipative. Solutions for the velocity and density profiles and their scaling with respect to $k_0$ are discussed in the supplementary material \cite{SM} and are shown in Figure~\ref{fig:disp}.  As there is an exact symmetry $\Omega(k) \to -\Omega(-k)$ due to the PT symmetry of the hydrodynamic equations with odd viscosity, we show only the $\Omega\geq 0$ part of the spectrum in Fig.~\ref{fig:disp}. The left panel of Figure~\ref{fig:disp} shows the bulk vorticity profile corresponding to the chiral boundary mode. We note here that the results for large $k/k_0$ are expected to be less universal as in applications they can be changed by higher order gradient corrections to the hydrodynamic equations.

\textit{\textbf{Discussion and conclusions.}}  
In this letter we considered a non-dissipative fluid with odd viscosity subject to the free surface boundary conditions. As the fluid dynamics is non-dissipative, the hydrodynamic equations can be obtained through the variational principle in Eqs.~(\ref{hydro-action}-\ref{action-edge}). The no-stress boundary condition is accounted for by the boundary action (\ref{action-edge}). The latter has a form of 1+1 action of two chiral boson fields $h$ and $\phi$ coupled to the background boundary geometry.  The field $h$ describes the form of the boundary and the field $\phi$ is an auxiliary field that can be removed for the price of making the boundary action nonlocal.

The Hamiltonian structure of the fluid dynamics derived for a fluid domain with no boundary allows for an interesting Hamiltonian reduction. Imposing the Hall constraint $Q=\nu^{-1} h/m$ makes all Casimirs (\ref{eq:IN}) proportional to the total mass of the fluid. The Hall constraint plays an important role in the hydrodynamics of the quantum Hall effect, and our interpretation makes this constraint very natural from the point of view of the Hamiltonian structure corresponding to the action (\ref{hydro-action}-\ref{action-edge}) \cite{SM}. 

Within the linearized dynamics, we showed that the velocity divergence in a compressible fluid with odd viscosity generates the vortical boundary layer at the surface of the fluid. The boundary layer is necessary to satisfy the tangent no-stress boundary condition. As a result, a chiral wave with the dispersion (\ref{dispersion}) can propagate along the fluid edge even in the absence of an external confining potential. The finite compressibility of the fluid regularizes singularities at the surface even in the absence of shear viscosity. 

The variational principle presented in this work is a good starting point for studying various issues related to surface excitations of fluids with broken parity. For example, one can use the action 
(\ref{hydro-action}-\ref{action-edge}) to obtain the nonlinear surface waves in the  incompressible limit by taking $c_s\rightarrow \infty$. In the lowest order in nonlinearity the result should match the action proposed in Ref.~\cite{abanov2018free} on phenomenological  grounds. 
Even more interestingly, one could impose the Hall constraint on the action (\ref{hydro-action}-\ref{action-edge}) and study the effects of odd viscosity on edge excitations of fluids similar to quantum Hall fluids. We reserve the investigation of these questions for future work.


\textit{\textbf{Acknowledgments.}} We are grateful to Paul Wiegmann for many discussions and for careful reading and commenting on an earlier draft of this work. AGA's research was supported by grants NSF DMR-1606591 and US DOE DESC-0017662. AGA and GMM are grateful to International Institute of Physics, Natal, Brazil for hospitality. GMM thanks Funda\c c\~ao de Amparo \`a Pesquisa do Estado de S\~ao Paulo (FAPESP) for financial support under grant 2016/13517-0. SG acknowledges support from PSC-CUNY Award.  SG acknowledges Aspen Center for Physics where part of this work was carried out, which is supported by National Science Foundation grant PHY-1607611.

\bibliographystyle{my-refs}
\bibliography{oddviscosity-bibliography.bib}

\begin{thebibliography}{10}
\bibitem{footnote-energy}
In this work we neglect all thermal and dissipative effects and energy
  conservation is a direct consequence of mass and momentum conservation.
  Indeed, it is easy to check that in the absence of external electric field
  the energy density $\frac{\rho v^{2}}{2}+\epsilon(\rho)$ with $p=\rho
  \epsilon_{\rho}-\epsilon$ is locally conserved.
\end{thebibliography}

 \newpage
 
\onecolumngrid
\bigskip

\section{Supplementary Information}

 \section{Bulk Variational Principle and Hamiltonian Structure}
 
In this section, we present a hydrodynamic action which provides the bulk equations (\ref{eq:eom}) for a compressible fluid with odd viscosity. For the brevity of notations we put the charge to mass ration $e/m\to 1$ and use the notation in Eq. (\ref{Clebsch}), namely, $u_\mu\equiv\p_\mu\theta +A_\mu +\alpha\p_\mu\gamma$ for $\mu=0,1,2$. Let us assume the fluid domain to be the whole two-dimensional plane. The generalization to the fluid domain with free surface will be discussed in the next section. Let us consider the action:
 \be
 	S[\theta, \alpha,\gamma,\rho,v_i]
	=-\int dt\int_{\mathbb R^2} d^2x\left[\rho (u_0+v^iu_i) \rho -\frac{\rho v^i v_i}{2}
	+\varepsilon(\rho)-\nu_ov_i\epsilon^{ij}\p_j\rho\right] \,. 
 \la{action-v}
\ee  

Variation of this action gives us
\begin{align}
	\delta S=\int dt\int &d^2x\left[\rho\,\delta v^i\left(v_i-u_i+\frac{\nu_o}{\rho}\p_i^*\rho\right)
	+(\delta\theta+\alpha\delta\gamma)\Big(\p_t\rho+\p_i(\rho v^i)\Big)
	+\rho\,\delta\gamma\left(\p_t\alpha+v^i\p_i\alpha\right)\right.
 \nonumber \\
	&\left.-(\rho\,\delta\alpha+\alpha\,\delta\rho)
	\left(\p_t\gamma+v^i\p_i\gamma\right)
	-\delta\rho\left(\p_t\theta+A_0+v^i(\p_i\theta+A_i)-\frac{1}{2}v^iv_i+\varepsilon'(\rho)
	-\nu_o\epsilon^{ij}\p_iv_j\right)\right]. 
\end{align}
From this variation, we obtain the Clebsch parametrization of velocity given in Eq.~(\ref{Clebsch}), the continuity equation and the dynamics of the Clebsch potentials:
\begin{align}
	& v_i = u_i -\nu_o \p_i^*\ln\rho\,,
 \\
	& \p_t\rho +\p_i(\rho v_i) =0\,,
 \\
 	& \p_t \alpha +v^i\p_i\alpha= 0\,,
 \\
 	& \p_t \gamma +v^i\p_i \gamma = 0\,
 \\
 	& \p_t\theta+A_0+v^i(\p_i\theta+A_i) = \frac{1}{2}v^iv_i-\varepsilon'(\rho)
	+\nu_o\epsilon^{ij}\p_iv_j \,.
\end{align}
Euler equation is obtained by taking the time derivative of $v_i$ from Eq.~(\ref{Clebsch}) and using the equations of motion to Clebsch potentials. Hence,
\begin{align}
	\p_t v_i&=\p_i\p_t\theta+\p_t\alpha\p_i\gamma+\alpha\p_i\p_t\gamma
	+\p_tA_i-\nu_o\p_i^*\p_t(\ln\rho) \,,
 \nonumber \\
	\p_tv_i&=\p_i\left[v^j\left(\frac{v_j}{2}-\p_j\theta-A_j\right)-\varepsilon'(\rho)
	+\nu_o\epsilon^{jk}\p_jv_k-A_0\right]-v^j\p_j\alpha\p_i\gamma-\alpha\p_i(v^j\p_j\gamma)
	+\p_tA_i+\nu_o\p_i^*\left[\frac{1}{\rho}\p_j(\rho v^j)\right] \,,
 \nonumber \\
	\p_tv_i&=-\p_i\left[v^j(\nu_o\p_j^*\ln\rho+\tfrac{1}{2}v_j)
	+\varepsilon'(\rho)-\nu_o\omega\right]+v^j(\p_i\alpha\p_j\gamma-\p_j\alpha\p_i\gamma)
	-E_i+\nu_ov^j\p_j\p_i^*\ln\rho+\nu_o\p_j\p_i^*v^j+\frac{\nu_o}{\rho}\p_i^*v^j\p_j\rho \,.
 \nonumber
\end{align}
In order to express the last line solely in terms of the fluid density and the velocity field, we must note that
\[
	v^j(\p_iv_j-\p_jv_i)
	=v^j(\p_i\alpha\p_j\gamma-\p_j\alpha\p_i\gamma)+B\epsilon_{ij}
	-\nu_ov^j(\p_i\p_j^*\ln\rho-\p_j\p_i^*\ln\rho) \,,
\]
and after a little bit of algebra, we end up with
\begin{align}
	&\p_tv_i=-v^j\p_jv_i-(B\epsilon_{ij}v^j+E_i)
	-\p_i\Big(\varepsilon'(\rho)-\nu_o\p_jv^{*j}\Big)
	+\frac{\nu_o}{\rho}\p_j\Big(\rho\,\p_i^*v^j\Big)-\frac{\nu_o}{\rho}\p_iv^{j}\p_j^*\rho\,,
 \nonumber \\
	&\p_tv_i+v^j\p_jv_i=-(B\epsilon_{ij}v^j+E_i)-\frac{1}{\rho}\p_j\left[p(\rho)\,\delta^j_i
	-\nu_o\rho\left(\p_i^*v^j+\p_iv^{*j}\right)\right] \,,
\end{align}
where we used the identity
\[
	\p_iv^j\p_j^*\rho=-\p_iv^{*j}\p_j\rho \,,
\]
together with fluid pressure definition, that is, $p(\rho)=\rho\varepsilon'(\rho)-\varepsilon(\rho)$.

\subsection{Hamiltonian Structure}

In the action (\ref{action-v}), the field $v_i$ is a Lagrange multiplier and can be ``integrated out''. Therefore, we can rewrite it as
\be
	S=-\int dt\int d^2x\left[\rho\left(\p_t\theta+A_0+\alpha\p_t\gamma+\frac{1}{2}v^iv_i\right)
	+\varepsilon(\rho)\right]. 
 \la{action-no-boundary}
\ee
Here the velocity field $v_i$ is expressed in terms of Clebsch parameters by Eq.~(\ref{Clebsch}). We separate time derivatives and rewrite (\ref{action-no-boundary}) as 
\be
	S=\int dt\left\{\int d^2x\Big[-\rho \p_t\theta-\rho\alpha\p_t\gamma \Big] - H\right\} \,, 
 \la{action-no-boundary2}
\ee
where the fluid Hamiltonian is given by
\be
	H=\int d^2x\left[\frac{1}{2}\rho v^iv_i+\varepsilon(\rho)+\rho A_0\right] \,. 
 \la{Hamiltonian}
\ee
The part of the action (\ref{action-no-boundary2}) containing time derivatives defines the Poisson algebra of the system. From the action (\ref{action-no-boundary2}), we see that $-\rho$ and $\theta$ are conjugated quantities whereas $-\rho\alpha$ is conjugated to $\gamma$. This means that we have the following Poisson brackets
\begin{align}
	\{\rho,\theta'\}&=-\delta(\boldsymbol x-\boldsymbol x'),
\\
	\{\rho\alpha,\gamma'\}&=-\delta(\boldsymbol x-\boldsymbol x'),
\\
	\{\theta,\gamma'\}&=0,
\\
	\{\rho,\rho'\alpha'\}&=0,
\\
	\{\rho,\gamma'\}&=0,
\\
	\{\rho\alpha,\theta'\}&=0.
\end{align}
Here, for the sake of brevity, we used the notations $\rho= \rho(t,\boldsymbol x)$, $\theta'=\theta(t,\boldsymbol x')$, etc.

It is straightforward to see that this canonical algebra is the same for fluids without odd viscosity. Namely, it is not hard to show that 
\begin{align}
	\{\rho,\rho'\}&=0 \,, 
 \la{rho-rho} \\
	\{\rho, u_i'\}&=\p_i\delta(\boldsymbol x-\boldsymbol x') \,, 
 \la{rho-u} \\
	\{u_i,u_j'\}&
	=\frac{\p_ju_i-\p_iu_j-(\p_jA_i-\p_iA_j)}{\rho}\delta(\boldsymbol x-\boldsymbol x')
	=\frac{\p_ju_i-\p_iu_j+\epsilon_{ij}B}{\rho}\delta(\boldsymbol x-\boldsymbol x') \,. 
 \la{u-u}
\end{align}
These brackets are the same as the brackets for density and velocity fields for a fluid without odd viscosity \cite{1997-ZakharovKuznetsov}. Therefore, the presence of odd viscosity leads to nothing but a redefinition of velocity field. One can easily obtain the Poisson brackets for $\rho$ and $v_i=u_i-\nu_o\p_i^*\ln\rho$ using (\ref{rho-rho}-\ref{u-u}). 

The algebra (\ref{rho-rho} - \ref{u-u}) is well-known and admits an infinite number of Casimirs, namely,
\be
	F=\int d^2x\,\rho\, f\left(\frac{\epsilon^{ij}\p_iu_j-B}{\rho}\right)
\ee
is a Casimir for any function $f$. Rewriting it in terms of vorticity $\omega=\epsilon^{ij}\p_iv_j$, we find 
\be
	F=\int d^2x\,\rho\, f(Q) \,,
\ee
where $Q$ is defined by Eq.~(\ref{eq:Q}). We can show that $\rho Q$ is a Casimir density by imposing that $f$ is a linear function of $Q$. 

From Eqs. (\ref{rho-rho}-\ref{u-u}), it is straightforward to work out the explicit form of Poisson algebra in terms of $\rho$ and $v_i$
\begin{align}
	\{\rho,\rho'\}&=0 \,, 
 \\
	\{\rho, v_i'\}&=\p_i\delta(\boldsymbol x-\boldsymbol x') \,, 
 \la{rho-v} \\
	\{v_i,v_j'\}&=-Q\epsilon_{ij}\,\delta(\boldsymbol x-\boldsymbol x')
	+\nu_o (\p_i\p_j'^*-\p_i^*\p_j')\frac{\delta(\boldsymbol x-\boldsymbol x')}{\rho} \,.
 \la{v-v}
\end{align}
Hamilton equations of motion for $\rho$ and $v_i$ follows from
\begin{align}
	\dot\rho&=\{H,\rho\},
 \\
	\dot v_i&=\{H,v_i\}+\p_tA_i \,,
\end{align}
where the fluid Hamiltonian is given by Eq.~(\ref{Hamiltonian}). Thus,
\be
	\dot\rho=\int d^2x'\, \rho'v'^i \{v_i',\rho\}=-\p_i(\rho v^i) \,,
\ee
and
\begin{align}
	\dot v_i&=\p_tA_i+\int d^2x' \left[\left(\frac{1}{2}v'^jv_j'
	+\frac{d\varepsilon'}{d\rho'}+A_0'\right)\{\rho', v_i\}+\rho'v'^j\{v_j', v_i\}\right]\,.
\end{align}
Substituting (\ref{rho-v}-\ref{v-v}) after some manipulations we obtain Euler equation corresponding to the stress (\ref{eq:Tij}):
\begin{align}
	\dot v_i&=-v^j\p_jv_i-(E_i+Bv_i^*)-\frac{1}{\rho}\p_j\left[p\,\delta^j_i
	-\nu_o\rho\left(\p_i^*v^j+\p_iv^{*j}\right)\right] \,.
\end{align}

\section{Variational Principle with Free Surface }

In this section, we generalize the bulk action (\ref{action-v}) to account for the free edge dynamics. For that, the hydrodynamic action must provide us the continuity and Euler equations together with kinematic and dynamic boundary conditions, Eq.~(\ref{eq:bcs}). Let us consider the case where the fluid domain is given by $y\leq h(t,x)$, thus the bulk action becomes
\be
 	S_{bulk}=-\int dt\int\limits_{-\infty}^{\infty} dx\int\limits_{-\infty}^{h(t,x)}dy 
	\left[\rho(u_0+v^iu_i)-\frac{\rho v^i v_i}{2}
	+\varepsilon(\rho)-\nu_ov_i\epsilon^{ij}\p_j\rho\right] \,. 
 \la{action-bulk2}
 \ee  
To vary this action, we must remember that the boundary function $h(t,x)$ is a dynamical field. Thus, using the Leibniz integral rule, we end up with
\begin{align}
	\delta S_{bulk}=\iint dt\, dx\int\limits_{-\infty}^{h(t,x)} &dy
	\left[\rho\,\delta v^i\left(v_i-u_i+\frac{\nu_o}{\rho}\p_i^*\rho\right)
	+(\delta\theta+\alpha\delta\gamma)\Big(\p_t\rho+\p_i(\rho v^i)\Big)
	+\rho\,\delta\gamma\left(\p_t\alpha+v^i\p_i\alpha\right)\right.
 \nonumber \\
	&\;\left.-(\rho\,\delta\alpha+\alpha\,\delta\rho)\left(\p_t\gamma+v^i\p_i\gamma\right)
	-\delta\rho\left(\p_t\theta+A_0+ v^i(\p_i\theta+A_i)-\frac{1}{2}v^iv_i+\varepsilon'(\rho)
	-\nu_o\epsilon^{ij}\p_iv_j\right)\right]
 \nonumber \\
	+\iint dt\,dx\Big\{\rho(\delta\theta&+\alpha\delta\gamma)[\p_th+v_x\p_xh-v_y]
	-\delta h\left[\rho\alpha(\p_t\gamma+v^i\p_i\gamma)+\varepsilon (\rho)
	-\nu_o\epsilon^{ij}v_i\p_j\rho+\rho(\p_t\theta+v^i\p_i\theta-\tfrac{1}{2}v^iv_i)\right]
 \nonumber \\
	+\,\nu_o&\delta\rho(v_x+v_y\p_xh)\Big\}_{y=h(t,x)} \,.
\end{align}
Variations of fields on the bulk are the same as in the previous section, hence they provide the same bulk equations. Thus, this action variation on bulk equations of motion becomes
\begin{align}
	\delta S_{bulk}\Big|_{on\; EoM} =&
	\iint dt\,dx\Big\{\rho(\delta\theta+\alpha\delta\gamma)[\p_th+v_x\p_xh-v_y]
	-\delta h\Big[p(\rho)-\nu_o\epsilon^{ij}\p_j(\rho v_j)\Big]
 \nonumber \\
	&+\,\nu_o\delta\rho(v_x+v_y\p_xh)\Big\}_{y=h(t,x)} \,.
 \la{bulk-bound-var}
\end{align}

Variations of $\theta$ and $\gamma$ on the edge give us the kinematic boundary condition, that is,
\be
	\p_th+v_x\Big|_{y=h}\p_xh-v_y\Big|_{y=h}=0\,.
 \la{eq:KBC20}
\ee
However, variation over $\rho$ on the boundary states that the tangent velocity vanishes at the edge, that is, $v_x+v_y\p_xh=0$ and variation over $h$ states that
\be
	\Big[p(\rho)-\nu_o\epsilon^{ij}\p_j(\rho v_j)\Big]_{y=h}=0 \,. 
 \la{h-bulk}
\ee

Obviously, these are not the no-stress boundary conditions from Eq.(\ref{eq:bcs}). Therefore, we must add purely boundary terms to the full action. Such boundary action can be described as
\begin{equation}
    S_{edge}=-\nu_o\iint dt dx\left[\tilde\rho \p_th\p_xh
    +\p_t\phi\p_x\phi-2\p_t\phi\sqrt{\tilde\rho\left(1+(\p_xh)^2\right)}\right] \,, 
 \la{edge-action}
\end{equation}
where we introduced the density boundary field $\tilde{\rho}(t,x)\equiv \rho\big(t,x,h(t,x)\big)$ and the field $\phi(t,x)$ as an independent boundary field (does not depend on $h$). To vary the edge action (\ref{edge-action}), we must take into account that the variations and derivatives of the boundary density $\tilde\rho$ are related to the boundary values of the variation of the bulk density $\rho$ in the following way:
\begin{align*}
	\delta\tilde\rho &=\delta\rho\big|_{y=h}+\delta h\,\p_y\rho\big|_{y=h} \,,
 \\
	\p_t\tilde\rho &=\p_t\rho\big|_{y=h}+\p_t h\,\p_y\rho\big|_{y=h} \,,
\\
	\p_x\tilde\rho &=\p_x\rho\big|_{y=h}+\p_x h\,\p_y\rho\big|_{y=h} \,.
\end{align*}
Hence,
\begin{align}
    & \delta S_{edge}=\nu_o\iint dt dx\left\{\delta\rho
    \left(\frac{\p_t\phi}{\sqrt{\rho}}\sqrt{1+(\p_xh)^2}-\p_th\p_xh\right)
    +2\delta\phi\left[\p_t\p_x\phi-\frac{\sqrt\rho\,\p_xh\,\p_x\p_th}{\sqrt{1+(\p_xh)^2}}
    -\frac{\p_t\rho+\p_th\p_y\rho}{2\sqrt\rho}\sqrt{1+(\p_xh)^2}\right]\right.
 \nonumber \\
     &+\left.\delta h\left[2\rho\,\p_x\p_th+\p_t\rho\p_xh+\p_th\p_x\rho 
     +\p_th\p_xh\p_y\rho-2(\p_x+\p_xh\p_y)\left(\frac{\p_t\phi\p_xh\,
     \sqrt{\rho}}{\sqrt{1+(\p_xh)^2}}\right)
     +\frac{\p_t\phi\p_y\rho}{\sqrt{\rho}}\sqrt{1+(\p_xh)^2}\right]\right\}_{y=h} \,.
 \la{sedge-var}
\end{align}

Combining the terms of $\delta\rho$ taken at the boundary from both bulk and boundary actions, 
we obtain that
\be
\left[v_x+v_y\p_xh-\p_xh\p_th+\frac{\p_t\phi}{\sqrt{\rho}}\sqrt{1+(\p_xh)^2}\right]_{y=h}=0.
\ee
Using the kinematic boundary condition, we can parametrize $\p_t\phi$ in terms of the hydrodynamic fields $(\rho, v_x, h)$.
\be
\p_t\phi=-\left[v_x\sqrt{\rho\big(1+(\p_xh)^2\big)}\right]_{y=h}. \la{eq:phi}
\ee

Plugging the kinematic boundary condition, Eq.~(\ref{eq:KBC20}), together with Eq.~(\ref{eq:phi}) into the equation of motion for $\phi$, we get that
\begin{align}
    &\left[-\big(\p_x+\p_xh\p_y\big)\left(v_x\sqrt{\rho\big(1+(\p_xh)^2\big)}\right)-\frac{\p_xh\,\p_x\p_th\sqrt\rho}{\sqrt{1+(\p_xh)^2}}-\frac{\p_t\rho+(v_y-v_x\p_xh)\p_y\rho}{2\sqrt\rho}\sqrt{1+(\p_xh)^2}\right]_{y=h}=0,\nonumber
    \\
    &\left[\frac{\p_t\rho+v_x\p_x\rho+v_y\p_y\rho}{2\sqrt\rho}+\sqrt{\rho}\big(\p_xv_x+\p_xh\p_yv_x\big)+\frac{v_x\p_xh\p_x^2h\sqrt{\rho}}{1+(\p_xh)^2}+\frac{\p_xh\sqrt\rho}{1+(\p_xh)^2}\big(\p_x+\p_xh\p_y\big)\Big(v_y-v_x\p_xh\Big)\right]_{y=h}=0, \nonumber
    \\
     &\left[-\frac{1+(\p_xh)^2}{2}(\p_xv_x+\p_yv_y)+\big(\p_xv_x+\p_xh\p_yv_x\big)+\p_xh\big(\p_xv_y+\p_xh\p_yv_y\big)\right]_{y=h}=0,\nonumber
     \\
     &\Bigg[\big(1-(\p_xh)^2\big)(\p_xv_x-\p_yv_y)+2\p_xh\big(\p_xv_y+\p_yv_x\big)\Bigg]_{y=h}=0. \la{Tns}
\end{align}

One can show with few lines of algebra that the term in the left hand side of Eq.~(\ref{Tns}) is proportional to the tangent component of the dynamic boundary condition, that is,
\be
 	n_is_jT^{ij}\Big|_{y=h}=0 \,,
 \la{eq:Tns10}
\ee
where $s_j$ are the components of the edge tangent vector.
 
Finally, let us turn our attention to the variation with respect to $h$. From the bulk action, we obtain the left hand side of Eq.~(\ref{h-bulk}). Combining (\ref{bulk-bound-var}) with (\ref{sedge-var}) and using Eq.(\ref{eq:phi}), we obtain for the total variation with respect to $h$:
\begin{align}
	&\left[\frac{p}{\nu_o}-\epsilon^{ij}\p_i(\rho v_j)
	+2\rho\,\p_x\p_th+\p_t\rho\p_xh+\p_th\p_x\rho 
	+\p_y\rho\left(\p_th\p_xh-v_x-v_x(\p_xh)^2\right)+2(\p_x+\p_xh\p_y)
	\left(\rho v_x\p_xh\right)\right]_{y=h}=0 \,,
 \nonumber \\
	&\left[\frac{p}{\nu_o}-\rho\,\p_xh(\p_xv_x+\p_yv_y)
	+\rho\,(\p_yv_x-\p_xv_y+2\p_x\p_th)
	-2\rho\, \p_xh(\p_xv_x+\p_xh\p_yv_x)-2\rho\,v_x\p_x^2h\right]_{y=h}=0 \,,
 \nonumber \\
	&\Bigg[p-\nu_o\rho\,(\p_yv_x+\p_xv_y)
	+\nu_o\rho\,\p_xh(\p_xv_x-\p_yv_y)\Bigg]_{y=h}
	=\sqrt{1+(\p_xh)^2}\;n^iT_{iy}\Big|_{y=h}=0 \,.
\end{align}
This condition together with (\ref{eq:Tns10}) implies that both component of the stress at the surface are vanishing $n_i T^{ij}=0$ and we recovered both dynamical boundary conditions.

To conclude this section, we showed that the action given by the sum of the bulk term (\ref{action-bulk2}) and of the boundary term (\ref{edge-action}) reproduces both bulk equations of motion and proper boundary conditions for the compressible two-dimensional fluid with odd viscosity and free surface. This action can be used to derive the Hamiltonian structure for the fluid with free surface. We will leave this problem for future.

\section{Linearized solutions}
In this section, we provide the derivation details of the linear surface wave dispersion and the corresponding density and velocity profiles. In the following, we express all wavevectors in units of $k_0=c_s/\nu_o$, and frequencies in units of $\Omega_0=\nu_o k_0^2$. Substituting the density and velocity (\ref{eq:twomodes}) into the the linearized dynamical boundary conditions (\ref{eq:112}) we obtain,
\begin{align}
	\sum_{\alpha=1}^2A_{\alpha}f_\alpha( k,\Omega)=0,\quad&
	\sum_{\alpha=1}^2A_{\alpha}g_\alpha( k,\Omega)=0
 \la{eq:disp10}\\	
	f_\alpha( k,\Omega)=\left(1+\frac{2 m_{\alpha} \Omega  k}{k^2- m^2_{\alpha}}
	-( m_\alpha^2+ k^2)\right),&\quad g_\alpha( k,\Omega)
	=\left(\Omega\frac{ k^2+ m_{\alpha}^2}{ k^2- m^2_\alpha}-2 m_\alpha  k\right)
 \la{eq:disp20}
\end{align}
These two equations define the eigenvalue problem for finding $\Omega$ and the corresponding amplitudes $A_{1,2}$. The decay rates $m_\alpha$ are subject to the condition $\operatorname{Re}(m_\alpha)>0$ and are given by the bulk dispersion relation Eq.~(\ref{eq:dispm}), that is $m_\alpha^2- k^2=\frac{1\pm \sqrt{1+4\Omega^2}}{2}$ for $\alpha=1,2$, respectively. It is important to know that the system of equations (\ref{eq:disp10}) and the relation (\ref{eq:dispm}) have exact PT symmetry $\Omega(k)=-\Omega(-k)$. The surface wave dispersion is found solving the compatibility condition of equations (\ref{eq:disp10}):
\begin{align}
	f_2( k,\Omega( k))g_1( k,\Omega( k))-f_1( k,\Omega( k))g_2( k,\Omega( k))=0 \,.
 \la{eq:SIdet}
\end{align}
The solution has PT symmetry $\Omega(k)=-\Omega(-k)$ and is plotted in Fig.~\ref{fig:disp}, for $\Omega>0$.

To obtain the solutions in two limits, i.e., $ k\ll 1$ and $ k\gg 1$, analytically, we write $\Omega(k)=\sum_j a_j  k^j$. Substituting this form into Eq.~\ref{eq:SIdet} and expanding it in the powers of $k$ we can find $a_j$ by setting the coefficients for each power to zero. The first three terms can be written as,
\begin{align}
 	\Omega &\approx-2\left( k^2- k^3+ k^4\right)+O( k^5), \qquad\quad \,\,\text{for} \,\, k\ll1\,,
 \la{eq:smallKdisp} \\
 	\Omega &\approx -\sqrt{2}  k+\frac{1}{4\sqrt{2}}\frac{1}{ k}
	+\frac{1}{8}\frac{1}{ k^2}+O( k^{-3}), \;\,\,\text{for} \,\, k\gg 1\,.
 \la{eq:largeKdisp}
\end{align}
For $k\approx 0.650551 $, another counter propagating surface mode emanates out of the bulk continuum  
\begin{align}
 	\Omega &\approx \sqrt{2}  k-\frac{1}{4\sqrt{2}}\frac{1}{ k}
	-\frac{1}{8}\frac{1}{ k^2}+O( k^{-3}), \; \,\,\text{for} \,\, k\geq 0.650551 \,.
\end{align}
Remarkably, the dispersion of this mode is just opposite in sign to the dispersion (\ref{eq:largeKdisp}) to all orders in the $1/k$ expansion and differs from it only due to non-perturbative corrections in, giving rise to the existence of the vortical boundary layer.

Let us now focus on the long wavelength limit $k\ll 1$. Using Eq.~(\ref{eq:smallKdisp}), we solve Eq.~(\ref{eq:disp10}) for the amplitudes $A_{1,2}$. Plugging it into Eq.~(\ref{eq:twomodes}) and restoring the dimensions, we end up with
\begin{align}
  	 n=-i \frac{A}{c_s} \left(2\frac{k}{ k_0} e^{k y}
	 -e^{ k_0 y}\right)e^{i  k x-i \Omega  t},
	 \quad v_x = i A \left(e^{k y}-e^{ k_0 y}\right)e^{i  k x-i \Omega  t}, 
	 \quad v_y = A \left(e^{k y}-\frac{k}{ k_0}e^{ k_0 y}\right)e^{i  k x-i \Omega  t} \,,
 \la{eq:profiles}
\end{align}
where $A$ is a free parameter defining the overall size of the wave. For completeness, we also give an expression for the height of the surface $h = -\frac{i A}{2\nu_o k^2} e^{ikx-i\Omega t}$ for $k\ll k_0$.

The vorticity corresponding to Eq.~(\ref{eq:profiles})  is given to the leading order by 
\begin{align}
	\omega\approx A  k_0 e^{ k_0 y}	e^{i  k x-i \Omega  t}\,.
 \la{eq:vorty=0}
\end{align}

Let us consider the incompressible limit of these linearized solutions. We fix the wave vector $k$ and the amplitude of the vertical velocity $v_y$ to be $A$. We then send $c_s\to \infty$ ($k_0\to \infty$). We find that the density is constant ($n\to 0$) and the velocities $v_x,v_y$ are finite in this limit everywhere near the boundary, while the vorticity (\ref{eq:vorty=0}) diverges near the surface $\omega \sim k_0\sim c_s$. Interestingly, the tangent velocity $v_x$ taken exactly at the surface $y=0$ vanishes in linear approximation and can have only values of higher order in the amplitude $A$ (beyond the linear approximation considered here). However, at finite depth of the order of $\delta=1/k_0\to 0$ the tangent velocity is finite and is of the order of $A$. Essentially, one can say that in the incompressible limit the tangent velocity has a finite discontinuity across the infinitesimal vortical boundary layer.

\begin{figure}
\centering
\includegraphics[scale=0.5]{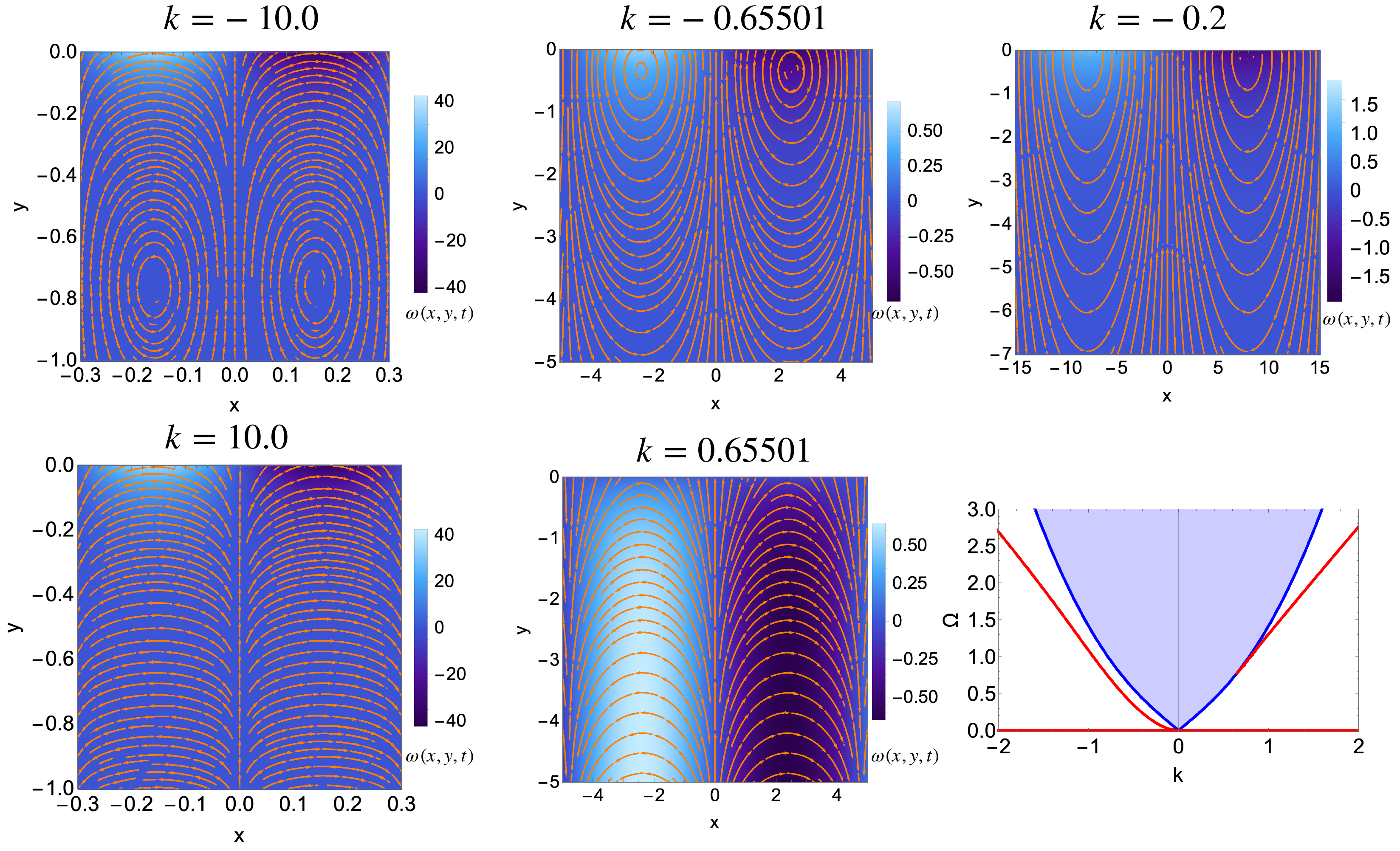}
\caption{Velocity streamlines overlaid over the vorticity profile are shown for different values of $k= \pm 10.0,  \pm 0.65501,  -0.2$. The lower right panel shows the dispersion of bulk (blue) and surface (red) waves. }
\label{fig:flow}
\end{figure}

We plot the linearized bulk velocity and vorticity profiles for different values of $k$ in Fig.~\ref{fig:flow}. The velocity profile given by the real parts of (\ref{eq:profiles}) are represented in the form of streamlines. The vorticity is plotted as color density plot in the background of the velocity streamline plots.  The odd viscosity dominates the flow for small negative $k$ shown in $k=-0.2$. The dimensionful surface dispersion for small negative $k$ is of the form $\Omega\sim -2\nu_o k|k| $. For intermediate and large negative values of $k$ the odd viscosity effects are suppressed and the dimensionful dispersion relation is of the form $\Omega\sim-\sqrt{2}c_s k$ which is independent of odd viscosity. Another evidence of suppression of parity breaking effects at large $k$ is shown by the emergence of a counter propagating mode for the $k\geq 0.65501$. For the  critical value of $k\approx 0.65501$, we see that the vorticity penetrates deep into the bulk due to the vanishing of the decay rate $m$ becoming a bulk mode. Since the bulk and boundary dispersion cross at this point, the disappearance of the boundary mode at low $k$ can be understood as due to a hybridization with the bulk.

In the $k\gg k_0$ limit the dispersion relation is of the form $\Omega=\pm\sqrt{2}c_sk$. The density and velocity profiles in this limit are of the form, 
\begin{align}
	n=\frac{ 2\sqrt{2} i A}{c_s}e^{ikx-i\Omega t} e^{|k| y}, \quad v_x=-2iA e^{ikx-i\Omega t} e^{|k| y}, \quad  v_y=2A e^{ikx-i\Omega t} e^{|k| y} 
\end{align}
The vorticity for this case is confined within the length $1/k \ll 1/k_0$ and is given by,
\begin{align}
	\omega=4iAk e^{ikx-i\Omega t} e^{|k|y}.
\end{align}
Even though the dispersion and the profile seems to be independent of odd viscosity $\nu_o$, the existence of a localized boundary mode is exclusively due to the presence of $\nu_o$. This is due to the fact that the tangent boundary condition resulting in the vortical boundary layer does not depend on the scale of the odd viscosity. However, a nonvanishing tangent boundary condition is solely a consequence of a non-zero odd viscosity.

\end{document}